\documentclass[11pt]{llncs}
\usepackage{makeidx}
\usepackage{graphicx}
 
\usepackage{ifthen}
\usepackage{microtype}
\usepackage{boxedminipage}
\usepackage{amsmath}
\usepackage{amsfonts}
\usepackage{amssymb}
\usepackage{authblk}
\usepackage{framed}

\usepackage{float}
\floatstyle{boxed}
\restylefloat{figure}

\newcommand{\DOM}{\{0, 1\}}

\newcommand{\TO}{\rightarrow}

\newcommand{\E}{{\bf E}}

\newcommand{\pr}{\mathbf{Pr}}

\newcommand{\veps}{{\varepsilon}}
\newcommand{\etal}{{\it et al.}}
\newcommand{\xor}{\oplus}

\begin{document}
 
\title{A note on the relation between XOR and Selective XOR Lemmas}
%
%
\author{Ragesh Jaiswal\thanks{Part of the work was done while the author was visiting University of California San Diego. \\ Email: \email{rjaiswal@cse.iitd.ac.in}}}
\authorrunning{Jaiswal et al.} 
%
%
\institute{Department of Computer Science and Engineering, \\
Indian Institute of Technology Delhi. 
}

\maketitle     

\begin{abstract}
Given an unpredictable Boolean function $f: \{0, 1\}^n \rightarrow \{0, 1\}$, the standard Yao's XOR lemma is a statement about the unpredictability of computing $\xor_{i \in [k]}f(x_i)$ given $x_1, ..., x_k \in \{0, 1\}^n$, whereas the Selective XOR lemma is a statement about the unpredictability of computing $\xor_{i \in S}f(x_i)$ given $x_1, ..., x_k \in \{0, 1\}^n$ and $S \subseteq \{1, ..., k\}$. 
We give a reduction from the Selective XOR lemma to the standard XOR lemma.
Our reduction gives better quantitative bounds for certain choice of parameters and does not require the assumption of being able to sample $(x, f(x))$ pairs.
\end{abstract}

\section{Introduction}
A boolean function $f: \DOM^* \TO \DOM$ is said to be $\delta(n)$-unpredictable for family of $s(n)$ size circuits iff for any any $n$ and any circuit $C$ of size $s(n)$, $\pr_{x \leftarrow \DOM^n}[C(x) \neq f(x)] \geq \delta(n)$, where $\delta: \mathbb{N} \TO \mathbb{R}$ and $s: \mathbb{N} \TO \mathbb{N}$ are functions over positive integers. Note that in the previous statement, the probability is over the uniform distribution on $\DOM^n$. The unpredictability can be defined more generally on a probability ensemble. 
However, in this work we work with the uniform distribution since the ideas developed here can easily be generalised.
Note that in case the circuit $C$ above is a randomised circuit, the probability is also over the internal randomness of the circuit.
The classical {\em Yao's XOR lemma} is a statement about unpredictability amplification by defining the XOR function that computes the xor of function values evaluated simultaneously at multiple inputs from $\DOM^n$.
More specifically, consider the function $F^{\xor k}(x_1, ..., x_k) \equiv \xor_{i} f(x_i)$, where $k$ is a positive integer and $x_1, ..., x_k \in \DOM^n$.
It can be shown that the function $F$ is $\veps(n)$-unpredictable for family of $s'(n)$ size circuits, where $\veps: \mathbb{N} \TO \mathbb{R}$ and $s': \mathbb{N} \TO \mathbb{N}$ are functions such that $\veps(n)$ is much smaller than $\delta(n)$ and $s'(n)$ is upper bounded by $s(n)$.
A theorem that is closely related to Yao's XOR lemma is the {\em direct product lemma}, that is a statement about the unpredictability of the direct product function $F^{|k}(x_1, ..., x_k) \equiv (f(x_1), ..., f(x_k))$. 
The XOR and direct product lemmas are connected and reductions between them are known~\cite{gnw95,u09,vw08,j08,ijkw10}.
One of the main ingredients in the reduction from the direct product to the XOR lemma is the {\em Selective XOR Lemma}.
The selective XOR lemma is a statement about the unpredictability of the function $F^k$ defined over strings of size $(nk+k)$ as $F^{k}(x_1, ..., x_k, r) \equiv \xor_{i} (f(x_i) \cdot r_i)$, where $x_1, ..., x_k \in \DOM^n$ and $r \in \DOM^k$.
In this work, we give a reduction from the selective XOR lemma to the standard XOR lemma.
Such reductions are already known. 
Goldreich~\cite{goldreich-book} gave a reduction that uses pairs $(x, f(x))$ for random $x \in \DOM^n$. 
Note that such pairs may be considered as non-uniformity  in a non-uniform computational setting.
Impagliazzo \etal~\cite{ijkw10} gave a reduction that does not require $(x, f(x))$ pairs.
We give a reduction that does not require $(x, f(x))$ pairs as in the reduction of Goldreich and our reduction gives better quantitative bounds on unpredictability than that of Impagliazzo \etal~\cite{ijkw10} for certain choice of parameters.
We discuss these in the related work subsection. 
We now give the formal statement of our reduction.

\begin{theorem}[Main Theorem]\label{thm:main}
If there is a (randomised) circuit $C$ of size $s$ such that 
$$\pr \left[C(x_1, ..., x_k) = F^{\xor k}(x_1, ..., x_k) \right] \geq \frac{1}{2} + \veps,$$
 then there is a randomised circuit $C'$ of size $\Omega(s)$ such that 
 $$\pr[C'(x_1, ..., x_k, r) = F^{k}(x_1, ..., x_k, r)] \geq \frac{1}{2} + \veps^2.$$
The probability in the above inequalities is over the internal randomness of the circuits and uniform choice of inputs (i.e., $x_1, ..., x_k \in \DOM^n$ and $r \in \DOM^k$).
\end{theorem}
We give the proof of this theorem in the next section. 
First, we look at the related work that gives reduction of the form given above.

\vspace{-0.1in}

\subsection{Related work}
Goldreich gave a reduction (Ex. 7.17 in \cite{goldreich-book}) that uses a circuit $C$ for computing $F^{\xor k}$ to construct a circuit $C'$ for $F^k$ in the following manner: Given an input $(x_1, ..., x_k, r)$ for $F^k$, the circuit $C'$ outputs $\left( C(x_1, ..., x_k) \xor \xor_{i: r_i=0} f(z_i) \right)$, where $z_1, ..., z_k$ are randomly chosen strings from $\DOM^n$. 
Note that for this reduction to work, we need $(z, f(z))$ pairs for randomly chosen $z \in \DOM^n$. 
In the non-uniform computational setting, this can be considered the {\em non-uniform advice} that is {\em hard-wired} to the circuit. 
Impagliazzo \etal~\cite{ijkw10} gave a reduction in the uniform computational model where availability of $(z, f(z))$ pairs is not required. 
They get around the $(z, f(z))$ pairs requirement by using the circuit $C$ that predicts $F^{\xor k}$ to construct a circuit $C'$ that predicts $F^{2k}$ (instead of $F^k$ as in the previous reduction).
On input $(x_1, ..., x_{2k}, r)$ the circuit $C'$ checks if $|\{i:r_i=1\}|$ is equal to $k$. 
If this is not the case, then the circuit $C'$ outputs a random bit. 
Otherwise, it constructs the input $(y_1, ..., y_k)$ using the $k$ strings in the set $\{x_i: r_i=1\}$,  and outputs $C(y_1, ..., y_k)$. 
Since $\Omega(\frac{1}{\sqrt{k}})$ fraction of the $2k$ bit strings have exactly $k$ 1's, it can be argued that if $\pr[C(x_1, ..., x_{k}) = F^{\xor k}(x_1, ..., x_k)] \geq \frac{1}{2} + \veps$, then $\pr[C'(x_1, ..., x_{2k}, r) = F^{2k}(x_1, ..., x_{2k}, r)] \geq \frac{1}{2} + \Omega(\frac{\veps}{\sqrt{k}})$.
Note that unlike the reduction of Goldreich, there is a factor of $\frac{1}{\sqrt{k}}$ in the advantage of the circuit $C'$ over $C$. 
In our reduction, there is a factor of $\veps$ instead of $\frac{1}{\sqrt{k}}$ as in Impagliazzo \etal's reduction. So, for certain parameters ($\veps \geq \frac{1}{\sqrt{k}}$) our reduction gives better quantitative bounds. 
Moreover, we give a reduction from the selective $k$-XOR lemma to the standard $k$-XOR lemma (instead of selective $2k$-XOR to standard $k$-XOR lemma).

\vspace{-0.1in}

\section{Proof of Theorem~\ref{thm:main}}
The inputs to the functions $F^{\xor k}$ and $F^k$ are assumed to be sets $\{x_1, ..., x_k\} \subseteq \DOM^n$ and not $k$-tuples.
This makes sense in the context of XOR lemmas since the order of inputs in a tuple is not important when taking XOR's.
However, we mention this explicitly since past literature use both $k$-tuples and $k$-sets when discussing XOR lemmas \footnote{Ideas from results using one input representation can easily be extended to the other. So, it is sufficient to discuss in terms of one input representation.}.
In the discussion below, we will use $w(r)$ to denote the number of 1's in the string $r \in \DOM^k$. 
Also, for a set $\{x_1, ..., x_k\}$, we use $\{x_1, ..., x_k\}_{|r}$ to denote the subset $\{x_i: r_i=1\}$.
Given a circuit $C$ such that $\pr[C(\{x_1, ..., x_k\}) = F^{\xor k}(\{x_1, ..., x_k\})] \geq \frac{1}{2} + \veps$, we construct the following circuit $C'$ for predicting $F^k$:

\begin{framed}
$C'(\{x_1, ..., x_k\}, r)$\\
\hspace*{0.2in} (1) \ \ \ If ($w(r)$ is odd) return a random bit\\
\hspace*{0.2in} (2) \ \ \ Randomly partition $\{x_1, ..., x_k\}_{|r}$ into sets $Y$ and $Z$ each containing $\frac{w(r)}{2}$ strings\\
\hspace*{0.2in} (3) \ \ \ Pick a random subset $S \subseteq \DOM^n \setminus (Y \cup Z)$ of size $(k-i)$.\\
\hspace*{0.2in} (4) \ \ \ Return $C(Y\cup S) \xor C(Z \cup S)$
\end{framed}

First, we show a lower bound on the conditional probability of the circuit $C'$ being correct given that $w(r) = 2i$ for any $i$.

\begin{lemma}
For any $1 \leq i \leq \lfloor \frac{k}{2}\rfloor$, $\pr[C'(\{x_1, ..., x_k\}, r)\ |\ w(r) = 2i] \geq \frac{1}{2} + 2 \veps^2$.
\end{lemma}
\begin{proof}
When given an input such that $w(r) = 2i$, the circuit partitions the set $\{x_1, ..., x_k\}_{|r}$ into $Y$ and $Z$, appends both sets with a randomly chosen set $S$ and then returns $C(Y\cup S) \xor C(Z \cup S)$.
The probability of the success of $C'$ on such inputs can be analysed using a bipartite graph described next.
Consider a bipartite graph $G = (L, R, E)$ where the vertices in $L$ correspond to subsets of $\DOM^n$ of size $(k-i)$ and vertices in $R$ correspond to subsets of size $k$. There is an edge from a subset $S$ of size $(k-i)$ in $L$ to a subset $T$ in $R$ iff $S \subset T$.
An edge $(S, T)$ is coloured green if $C(T) = F^{\xor k}(T)$ and is coloured red otherwise (i.e., green indicates that $C$ gives correct answer on $T$).
For any vertex $S$ on the left, let $\gamma_S$ denote the fraction of edges incident on it that are green. We know that $\E[\gamma_S] \geq \frac{1}{2} + \veps$ since this is the probability with which circuit $C$ succeeds.
The probability of success of circuit $C'$ is the probability that for a randomly chosen vertex on the left and two of its random edges, either both the edges are green or both are red. This probability is given by:
\[\E \left[ \gamma_S^2 + \left(1 - \gamma_S \right)^2 \right] = 1 - 2 \E[\gamma_S] + 2 \cdot \E[\gamma_S^2] \geq \frac{1}{2} + 2 \veps^2.\]
The last inequality follows from Cauchy-Schwarz.\qed
\end{proof}
Analysing the success probability of $C'$ is now simple using the above lemma. The circuit $C'$ outputs a random bit when $w(r)$ is odd. So the conditional probability of success given that $w(r)$ is odd is $\frac{1}{2}$. The above lemma suggests that the conditional probability of success of $C'$ given that $w(r)$ is even is $\geq \frac{1}{2} + 2 \cdot \veps^2$. Combining these, we get that the probability of success of $C'$ is at least $\frac{1}{2} + \veps^2$. This concludes the proof of Theorem~\ref{thm:main}.

\vspace{0.1in}

\noindent
{\bf Acknowledgements} The author thanks Oded Goldreich for useful comments on a earlier version.

\bibliographystyle{plain}
\bibliography{biblio}

\end{document}